\input amstex
\documentstyle{amsppt}
\magnification=\magstephalf
 \addto\tenpoint{\baselineskip 15pt
  \abovedisplayskip18pt plus4.5pt minus9pt
  \belowdisplayskip\abovedisplayskip
  \abovedisplayshortskip0pt plus4.5pt
  \belowdisplayshortskip10.5pt plus4.5pt minus6pt}\tenpoint
\pagewidth{6.5truein} \pageheight{8.9truein}
\subheadskip\bigskipamount
\belowheadskip\bigskipamount
\aboveheadskip=3\bigskipamount
\catcode`\@=11
\def\output@{\shipout\vbox{%
 \ifrunheads@ \makeheadline \pagebody
       \else \pagebody \fi \makefootline 
 }%
 \advancepageno \ifnum\outputpenalty>-\@MM\else\dosupereject\fi}
\outer\def\subhead#1\endsubhead{\par\penaltyandskip@{-100}\subheadskip
  \noindent{\subheadfont@\ignorespaces#1\unskip\endgraf}\removelastskip
  \nobreak\medskip\noindent}
\outer\def\enddocument{\par
  \add@missing\endRefs
  \add@missing\endroster \add@missing\endproclaim
  \add@missing\enddefinition
  \add@missing\enddemo \add@missing\endremark \add@missing\endexample
 \ifmonograph@ 
 \else
 \vfill
 \nobreak
 \thetranslator@
 \count@\z@ \loop\ifnum\count@<\addresscount@\advance\count@\@ne
 \csname address\number\count@\endcsname
 \csname email\number\count@\endcsname
 \repeat
\fi
 \supereject\end}
\catcode`\@=\active
\CenteredTagsOnSplits
\NoBlackBoxes
\nologo
\def\today{\ifcase\month\or
 January\or February\or March\or April\or May\or June\or
 July\or August\or September\or October\or November\or December\fi
 \space\number\day, \number\year}
\define\({\left(}
\define\){\right)}
\define\Ahat{{\hat A}}

\define\End{\operatorname{End}}

\define\HH{{\Bbb H}}

\define\RR{{\Bbb R}}

\define\ZZ{{\Bbb Z}}
\define\[{\left[}
\define\]{\right]}
\define\ch{\operatorname{ch}}
\define\chiup{\raise.5ex\hbox{$\chi$}}

\define\exertag #1#2{#2\ #1}

\define\protag#1 #2{#2\ #1}

\define\res#1{\negmedspace\bigm|_{#1}}
\define\temsquare{\raise3.5pt\hbox{\boxed{ }}}

\define\theprotag#1 #2{#2~#1}

\define\xca#1{\removelastskip\medskip\noindent{\smc%
#1\unskip.}\enspace\ignorespaces }

\define\zmod#1{\ZZ/#1\ZZ}

\define\zt{\zmod2}

\NoRunningHeads 

\define\Det{\operatorname{Det}}
\define\Ebar{\overline{E}}
\define\Euler{\operatorname{Euler}}
\define\KK{\hat K}
\define\KO{\widehat{KO}}
\define\KSp{\widehat{KSp}}
\define\Pfaff{\operatorname{Pfaff}}
\define\Qbar{\overline{Q}}
\define\one{\bold{1}}
\define\scrK{\Cal K}
\redefine\HH{\hat H}

\refstyle{A}
\widestnumber\key{SSSSSSSSSS}   
\document

	\topmatter
 \title\nofrills On Ramond-Ramond Fields and $K$-Theory \endtitle
 \author Daniel S. Freed\\Michael Hopkins  \endauthor
 \thanks The first author is supported by NSF grant DMS-9626698.  The second
author is supported by NSF grant DMS-9803428.\endthanks
 \affil Department of Mathematics, University of Texas at Austin\\ 
	Department of Mathematics, Massachusetts Institute of Technology
	\endaffil  
 \address Department of Mathematics, University of Texas, Austin, TX
78712\endaddress 
 \email dafr\@math.utexas.edu \endemail
 \address Department of Mathematics, Massachusetts Institute of Technology,
Cambridge, MA 02139\endaddress 
 \email mjh\@math.mit.edu \endemail
 \date May 17, 2000\enddate
 \abstract 
 A recent paper by Moore and Witten~\cite{MW} explained that Ramond-Ramond
fields in Type~II superstring theory have a global meaning in $K$-theory.  In
this note we amplify and generalize some points raised in that paper.  In
particular, we express the coupling of the Ramond-Ramond fields to D-branes
in a $K$-theoretic framework and show that the anomaly in this coupling
exactly cancels the anomaly from the fermions on the brane, both in Type~IIA
and Type~IIB.
 \endabstract
	\endtopmatter

\document

 \comment
 lasteqno @ 15
 \endcomment

The proper quantization condition for the Ramond-Ramond (RR) field strengths
in Type~II superstring theory has been the subject of numerous studies, most
recently a paper by Moore and Witten~\cite{MW}.  They assert that RR~fields
have a characteristic class in integral $K$-theory, in the same sense that a
1-form field which globally is a $U(1)$~connection has a first Chern class in
second integral cohomology.  They also define the partition function of these
fields, based on~\cite{W1}, and discuss the coupling to D-branes in its
standard expression with differential forms and the resulting anomaly (in
Type~IIA).  Our main goal is to explain that this coupling is most naturally
expressed in the $K$-theoretic framework (equation~\thetag{15} below) and that
the coupling term has an anomaly which cancels the anomaly from fermions on
the D-brane.  This anomaly cancellation does not involve the RR~partition
function or the quadratic form needed to define it.  It works in both
Type~IIA and Type~IIB, though the details depend on the dimension of the
D-brane.  The cancellation holds for local and global anomalies.

We begin with a baby example from ordinary electromagnetism in four
dimensions which we find useful in thinking about self-dual fields (better:
fields with self-dual field strength).  Then we review standard material
about electric and magnetic coupling for $p$-form fields.  For self-dual
$p$-forms we make the simple observation that an electrically charged object
is also magnetically charged and {\it visa versa\/}.  Next, we hint at the
correct mathematical framework which mixes integral $K$-theory and
differential forms---a differential-geometric form of $K$-theory, which we
call {\it differential $K$-theory\/}---so is the proper home for RR~fields.
This is the subject of a forthcoming joint paper with I. Singer.  In the
presence of D-branes, because of the magnetic charge there is a shift in the
meaning of the RR~fields; some of the details are dictated by the anomaly
cancellation.  Finally, we express the electric coupling of RR~fields and
D-branes in geometric $K$-theory and compute the anomaly.\footnote{We only
sketch the argument here and leave a detailed proof for the subsequent
paper.}  The anomaly cancellation uses a geometric form of the Atiyah-Singer
index theorem for families of real Dirac operators~\cite{AS}.

Our entire discussion assumes that the $B$-field of Type~II vanishes.
 
We are not careful with factors of~$2\pi $; the qualitative ideas discussed
here do not depend on them and in any case they depend on conventions.  On
the other hand, certain factors of~$1/2$ play a crucial role.  A paper of
Cheung and Yin~\cite{CY} was instrumental in our understanding of these
factors.

Cheung and Yin also demonstrate the perturbative anomaly cancellation for
Type~IIB D-branes.\footnote{At the end of~\S4 in~\cite{CY} there remains a
puzzle about the D3-brane in Type~IIB.  Our approach to the anomaly resolves
it.  In fact, since the Ramond-Ramond field in our formulation has self-dual
field strength, {\it all\/} D-branes are both electrically and magnetically
charged.}  Their work refines an earlier paper of Green, Harvey, and
Moore~\cite{GHM}.  A simultaneous paper of Moore and Minasian~\cite{MM} also
discusses the perturbative anomaly cancellation.  They go further in that
they also suggest the connection of D-brane charge to $K$-theory.  The global
anomaly was not treated in these references, and in~\cite{MW} it was only
discussed in a particular example.

Our lagrangians often include both a field and its dual, or self-dual fields.
In Lorentzian classical field theory, for example on Minkowski spacetime,
self-duality may be imposed as an external constraint not derived from an
action principle.  The lagrangian gives Poisson brackets and one can proceed
with canonical quantization in this framework.  In Euclidean quantum field
theory, where one computes partition functions and correlation functions
using the functional integral, the self-duality constraint is not imposed on
the classical fields but rather one imposes it in defining the functional
integral.  In this paper we work in Euclidean field theory, so do not impose
the self-duality constraint on the classical fields; as we do not integrate
over those fields we never encounter the subtleties of their quantization.

Anomaly cancellation is not sufficient to define correlation functions.
Geometrically, the exponentiated (effective) action is naturally a
section~$s$ of a hermitian line bundle with connection over a space~$S$ of
fields, and correlation functions are integrals over~$S$.  The absence of
anomalies is the assertion that the line bundle admits a flat section~$\one$
of unit norm, and then one takes the exponentiated action to be the
ratio~$s/\one$.  Now $\one$~is unique up to a phase on each connected
component of~$S$.  The overall phase is irrelevant, but relative phases are
important.\footnote{They arise in many contexts, often as ``$\theta
$-angles''.  As with other ingredients in field theory, they are constrained
by locality.}  In our case the existence of an isomorphism between the line
bundle for the fermion pfaffian and the (inverse) line bundle for the
coupling of the RR~fields and the D-brane is the index theorem.  It appears
that the index theorem can be strengthened to a {\it canonical\/}
isomorphism.  This would fix a trivializing section~$\one$ and eliminate
potential ambiguities in the definition of this part of the effective action.
We hope to resolve this issue in a future paper.

We thank Robbert Dijkgraaf, Jacques Distler, Greg Moore, Graeme Segal, Is
Singer, and Ed Witten for extensive conversations and correspondence on many
of the topics discussed here.

 \subhead A baby example
 \endsubhead

Consider Maxwellian electromagnetism on an oriented Riemannian
4-manifold~$X$.  A gauge field~$A$ is locally a 1-form and globally a
$U(1)$~connection.  Its curvature~$F$ satisfies the Bianchi identity $dF=0$
and the field equations in empty space assert $d*F=0$.  There is a dual gauge
field~$A'$ whose curvature~$F'$ satisfies
  $$ F'=*F. \tag{1} $$
In the usual formulation we choose either~$A$ or~$A'$ as the fundamental
field and write everything in terms of it.  Electrically charged objects with
respect to~$A$ are magnetically charged with respect to~$A'$ and {\it visa
versa\/}. 
 
As a toy model for self-dual fields, consider a formulation which includes
both~$A$ and~$A'$ as fundamental fields.  Thus the pair~$(A,A')$ functions as
a self-dual field.  (At the end of the introduction we remarked on field
theories with self-dual fields.)  Notice that the characteristic class
of~$(A,A')$ is an element of the group~$\Gamma(X) :=H^2(X)\oplus H^2(X)$.  To
define the quantum theory\footnote{We include these remarks, which we found
instructive, even though we do not quantize self-dual gauge fields in our
application to Type~II superstring theory.} we need, as explained
in~\cite{W2}, a symplectic form on~$\Gamma (X)$ and a quadratic refinement of
its reduction modulo two.  Assume $X$~is compact.  The symplectic form is
  $$ \omega \bigl((x,x'),(y,y') \bigr) = x\cdot y' - y\cdot x' \tag{2} $$
and the quadratic refinement is 
  $$ Q(x,x') = x\cdot x' \pmod2. \tag{3} $$
In these expressions the dot is the cup product pairing followed by
evaluation on the fundamental class.  The two natural polarizations
of~$\Gamma (X)$ lead to the description of the partition function in terms
of~$A$ or~$A'$; the form~$Q$ vanishes in these cases. 
 
E. Witten pointed out that $Q$~is preserved by the $SL(2;\ZZ)$~action only if
the intersection pairing is even, that is, only if $X$~is spin.  Thus the
same is true of the partition function.  This result is explained a different
way in~\cite{W3}.
 
The standard Euclidean kinetic term for a gauge field~$A$ is~$\frac 12|F|^2$,
which in local coordinates is~$\frac 14F_{\mu \nu }F_{\mu '\nu '}g^{\mu \mu
'}g^{\nu \nu '}$.  (As usual $g$~is the Riemannian metric.)  In the $(A,A')$
action there are kinetic terms for both~$A$ and~$A'$, so each appears in the
action as $1/2$~its usual value:
  $$ \int_{X}\left\{ \frac 14|F|^2 + \frac 14|F'|^2 \right\}\,\text{vol}_X.
     \tag{4} $$
The extra factor of~$1/2$ is perhaps clearest in Minkowski spacetime, where
the constraint~\thetag{1} is imposed on the Euler-Lagrange equation.

 \subhead Electric and magnetic charge
 \endsubhead

Continuing for the moment with the ordinary Maxwell theory of a single gauge
field~$A$, a particle has a worldline~$W\subset X$, an oriented compact
one-dimensional submanifold of~$X$ (not necessarily connected).  Suppose the
particle is electrically charged with respect to~$A$.  This is implemented by
adding an interaction term to the Euclidean lagrangian:
  $$ i\int_{W} qA. \tag{5} $$
Here $q$~is a locally constant function on~$W$ which represents the number of
units of charge carried by the particle.  Charge quantization of the quantum
theory asserts that $q$~is integral (in appropriate units).  Strictly
speaking, only the exponential of \thetag{5} is well-defined; it is
interpreted as a power of the holonomy of the connection~$A$.  Topologically,
$q$~determines an element of~$H^0(W)$.  The contribution of the particle to
the total electric charge on~$X$ is the pushforward of~$q$ in cohomology by
the inclusion of~$W$ into~$X$, which is a map $H^0(W)\to H^3_c(X)$.  Here we
take the image to be in {\it compactly supported\/} cohomology.  In
particular, the topological class of electric charge is an element
of~$H^3_c(X)$.  At the level of differential forms, the electric charge is
represented by a 3-form~$j=j_{(W,q)}$ with compact support, the Noether
current, and the field equation is modified to $d*F = j$.  Strictly speaking,
$j$~is canonically defined only as a distribution supported on~$W$, but as we
will see it is often important to smooth it out.  As explained in~\cite{MW},
the electric charge lies in the kernel of the natural map $H^3_c(X)\to
H^3(X)$.  Over~$\RR$ this follows from the field equation $d(*F)=j$.
 
Magnetic charge has a quite different classical description, and is most
easily illustrated for a 0-form gauge field~$C$ in a three-dimensional
spacetime~$Y$ (which we again take to be compact oriented Riemannian).  In
other words, $C$~is a circle-valued scalar field on~$Y$.  A particle may be
magnetically charged with respect to~$C$, so suppose $W\subset Y$ is a
compact oriented one-dimensional submanifold, the worldline of a particle.
As before, let $q$~be a locally constant integer-valued function on~$W$.
Then to say the particle is magnetically charged with respect to~$C$ shifts
the meaning of~$C$.  Usually one says that $C$~is defined only on the
complement of~$W$ and has winding number~$q$ around~$W$.  However, for many
purposes it is not sufficient to have $C$~defined on a subset of spacetime.
In particular, it would not make sense then to restrict~$C$ to~$W$, something
we need to do in the superstring theory context later on.  Hence we need to
explain in what geometric sense $C$~extends over~$W$.  Obviously, $C$~does
not extend as a function over~$W$.  Instead, we represent the image of~$[q]$
under the pushforward $H^0(W)\to H^2_c(Y)$ as the first Chern class of a
circle bundle~$P=P_{(W,q)}$ with smooth connection.  Then $C$~extends to all
of~$Y$ not as a circle-valued function, but rather is a section of~$P$.  The
field strength of~$C$ is now the ``covariant derivative'' of the section,
more precisely the connection form relative to the section~$C$.  It is not
closed; its differential is the curvature of~$P$. Furthermore, as part of the
construction of the ``geometric magnetic current''~$P$ we may fix a
trivialization on the complement of~$W$.  Then the ratio of~$C$ to that fixed
trivialization, defined on the complement of~$W$, is the usual description of
a circle-valued gauge field with winding number~$q$ about~$W$.  Notice that
although the Chern class of~$P$ is canonically determined from~$(W,q)$, the
connection and curvature are not.\footnote{We can, however, use the metric
on~$X$ and a cutoff function to construct a {\it smooth\/} curvature form and
also a connection with this curvature (as well as a trivialization on the
complement of~$W$).  This leads to a coupling between the metric on~$X$ and
magnetic charge which was the crucial idea in understanding the anomaly
cancellation for the $M$-theory 5-brane~\cite{FHMM}.}  Returning to Maxwell
theory in four dimensions, a magnetically charged particle worldline~$W$ with
locally constant function~$q$ gives rise to a class in~$H^3_c(X)$, which is
now represented by a {\it gerbe\/} with connection, and the meaning of the
gauge field~$A$ is now shifted in an analogous way by this gerbe.  The
connection on the gerbe is represented locally by a 2-form.  (We discuss
$p$-form fields in general in the next section.)  As with electric charge,
the topological class of magnetic charge is an element in the kernel of
$H^3_c(X)\to H^3(X)$.  The existence of~$A$ is a geometric form of the
assertion that the image of magnetic charge in~$H^3(X)$ vanishes.  The
curvature~$j$ of the gerbe is a 3-form, and at the level of differential
forms the Bianchi identity is modified to~$dF=j$.

If $W$~is electrically charged with respect to~$A$, then it is magnetically
charged with respect to the dual gauge field~$A'$.  At the level of
differential forms this follows from equation~\thetag{1} and the last
equations in each of the two preceding paragraphs.  A more precise argument
including the geometric form of magnetic charge, at least in dimensions two
and three, is given in~\cite{W5, Lecture~8}.
 
Next, we simply observe that objects charged with respect to a self-dual
gauge field carry electric and magnetic charge simultaneously.  This is clear
from the previous discussion for our toy model~$(A,A')$: a particle
electrically charged under~$A$ is magnetically charged under~$A'$, so both
electrically and magnetically charged under~$(A,A')$.  An important subtlety
here is the normalization of the coupling term.  Since the kinetic
terms~\thetag{4} in the $(A,A')$~action appear with $1/2$~the usual
coefficient, so does the interaction term~\thetag{5}.  The entire action is
  $$ \int_{X}\left\{ \frac 14|F|^2 + \frac 14|F'|^2
     \right\}\,\text{vol}_X\,+\,i\int_{W}\frac 12\,qA. \tag{6} $$
(Recall that in the classical theory we impose the equation $F'=*F$ in the
$(A,A')$~formulation.)  One easy way to see the~$1/2$ in the coupling term is
to compute the Euler-Lagrange equation from varying~$A$---an overall factor
in a classical lagrangian does not affect the equations of motion.  We can
also verify that in the quantum theory elimination of~$A'$ leads to the
standard action for a charged particle.  Namely, in~\thetag{6} the particle
is magnetically charged with respect to~$A'$, which is implemented by a shift
in the geometric meaning of~$A'$.  Under duality that shift goes over into
the electric coupling of~$A$ to the particle, and since the kinetic term
for~$A'$ has $1/2$~its usual coefficient, so does this coupling term under
duality.  Added to the coupling term already present in~\thetag{6} we recover
the standard action upon eliminating~$A'$, as claimed.
 
The factor of~$1/2$ in the coupling term is at first sight problematical,
since for~$q=1$ the exponentiated action seemingly involves a square root of
holonomy, which is anomalous.  However, upon closer examination we find that
in the self-dual formulation the periods of~$F,F'$ are {\it even\/}, and this
extra factor of~2 renders the exponentiated action well-defined.  (One can
see this factor of~2 in the quantization of the self-dual field using
$\vartheta $-functions, for example.)

As another example, consider a theory on a six-dimensional manifold~$X$ which
includes a chiral 2-form gauge field~$B$.  If a 1-brane $W\subset X$, which
is a closed oriented 2-dimensional submanifold, is electrically charged with
respect to~$B$, then it is also magnetically charged.  Then we have two
effects simultaneously: the meaning of~$B$ is shifted---in particular, $H$~is
not closed but rather its differential is a closed 4-form Poincar\'e dual
to~$W$---{\it and\/} there is a term
  $$ i\int_{W}\frac 12\,qB \tag{7} $$
included in the action.\footnote{As in the self-dual $(A,A')$ theory above,
the quantization law for~$B$ involves a factor of~2 which makes~\thetag{7}
well-defined.  More precisely, the de Rham cohomology class of the field
strength of~$B$ lies in the image of~$H^3(X;\ZZ)$ in~$H^3(X;\RR)$ by {\it
twice\/} the usual map.}  Observe that \thetag{7}~is of ``Green-Schwarz
type''.  In particular, it has an anomaly---its exponential is not a unit
norm complex number, but rather is a unit norm element in an abstract complex
line.  Furthermore, if we study~\thetag{7} as a function of parameters, we
obtain a complex line bundle over the parameter space with metric and
connection, and then \thetag{7}~is a unit norm section which is not
necessarily covariant constant.  The line bundle is topologically trivial,
but it may have nonzero curvature and/or holonomy.  Such terms in actions
contribute to the overall anomaly, which in geometric form is the curvature
and holonomy of the tensor product of all such line bundles arising in the
exponentiated action.  Physicists usually express the anomaly by saying that
terms like~\thetag{7} are not invariant under gauge transformations (of the
$B$~field).

 \subhead Local fields and global topology
 \endsubhead

It is well-known that fields which are locally represented by a differential
$p$-form often have a global significance which is more intricate than a de
Rham cohomology class.  The first author initially encountered this in
Chern-Simons theory~\cite{F1};\footnote{Embarrassingly, the two papers
promised in~\cite{F1} never appeared---there was trouble finding the proper
context in which to work out the details of integration for global $p$-form
fields of this type.  The work of Hopkins and Singer cited below, as well as
our forthcoming work mentioned in the introduction, fills this gap.} in the
physics literature it appears earlier in this context~\cite{A}, \cite{G} and
implicitly even earlier in supergravity theories.  The prototype is a 1-form
gauge field which globally is a $U(1)$~bundle with connection.  Up to
equivalence the $p$-form analog for higher~$p$ is a class in smooth Deligne
cohomology~\cite{D}, \cite{B} or equivalently a Cheeger-Simons differential
character~\cite{CS}.  But it is important in the physics that fields be
defined on the nose, not simply up to equivalence.  This is required by
locality, for example, and is also necessary if one is to speak properly
about group actions.  Put differently, we need a framework which includes
automorphisms of these fields.  Furthermore, one also encounters
``trivializations'' or ``sections'' of this type of $p$-form field, as we did
in the our discussion of magnetic charge.\footnote{Another important example
in physics is the $B$-field in Type~I supergravity coupled to super
Yang-Mills---even in the classical theory without the Green-Schwarz term.}  A
mathematical treatment including these refinements appears in ongoing work of
the second author with I.~Singer~\cite{HS}.\footnote{For an outline of what
is needed, see also~\cite{DF,\S6.3}.}  For our purposes here we simply remind
that the set of equivalence classes of $p$-form fields analogous to
$U(1)$~connections on a manifold~$X$, denoted\footnote{We use the grading
most natural in Deligne cohomology.}~$\HH^{p+1}(X)$, fits into the exact
sequence
  $$ 0 @>>> H^p(X;\RR)/H^p(X;\ZZ) @>>> \HH^{p+1}(X) @>>> A^{p+1}(X) @>>> 0,
     \tag{8} $$
where 
  $$ A^{q}(X) := \{(\lambda ,\omega )\in H^q(X;\ZZ)\times \Omega
     ^q_{\text{closed}}(X): \lambda _{\RR}=[\omega ]_{\text{de
     Rham}}\}. \tag{9} $$
One should think of~$\lambda $ as the characteristic class and~$\omega $ as
the curvature, or field strength.  The first term is the torus of
topologically trivial flat elements.  (A more detailed heuristic exposition
of this type of $p$-form fields appears in~\cite{FW,\S6}.)
 
There is a version of this story for any generalized cohomology theory.  The
generalized cohomology maps onto a full lattice in ordinary real cohomology.
In physical applications of these ideas there are Dirac quantization
conditions which dictate the proper lattice, and we choose a generalized
cohomology theory accordingly.\footnote{The choice of map to real cohomology
is also part of the Dirac quantization condition.  For example, we noted
above a factor of~2 in this map for certain self-dual fields.}  The choice
appropriate to RR~fields in Type~II is $K$-theory.  Then the basic exact
sequence for equivalence classes~$\KK^{p+1}(X)$ is
  $$ 0 @>>> K^p(X;\RR)/K^p(X) @>>> \KK^{p+1}(X) @>>> B^{p+1}(X) @>>> 0,
     \tag{10} $$
where 
  $$ B^{q}(X) := \{(x ,\omega )\in K^q(X)\times \Omega
     ^q_{\text{closed}}(X): \ch(x)=[\omega ]_{\text{de
     Rham}}\}. \tag{11} $$
By Bott periodicity only the parity of~$p$ matters.  There is a
category~$\scrK^{p+1}(X)$ whose equivalence classes are elements
of~$\KK^{p+1}(X)$.

The Ramond-Ramond fields of Type~II superstring theory (with vanishing
$B$-field), taken together in all degrees, are objects in~$\scrK^q(X)$.  (The
parity of~$q$ is even for Type~IIA and odd for Type~IIB.  The self-duality
condition on the field strength is imposed when quantizing these fields,
which we do not do in this paper.)  This is the precise version of the
proposal made in~\cite{MW}, particularly their equation~\thetag{2.17}, which
asserts that (up to a factor of~$2\pi $) the field strength~$G$ of an
RR~field which maps to~$(x,\omega )\in B^{\bullet}(X)$ is\footnote{Here we
interpret~$\sqrt{\Ahat(X)}$ as a differential form.  Note that it is
invertible (of the form~$1+\text{nilpotent}$).  One explanation of its
appearance is the following.  On a compact spin$^c$ manifold~$X$, both
$K^{\bullet}(X)\otimes \RR$ and~$H^{\bullet}(X;\RR)$ carry an addition,
multiplication, and a bilinear form.  The Chern character preserves the
addition and multiplication but not the bilinear forms.  The modification
by~$\sqrt{\Ahat(X)}$, as in~\thetag{12}, preserves addition and the bilinear
forms but not the multiplication.  The physics uses the addition
(superposition of states in quantum mechanics) and the bilinear form (for
example, in the coupling term~\thetag{14} below), but not as far as we can
tell the multiplication.}
  $$ G = \sqrt{\Ahat(X)}\;\omega. \tag{12} $$
Several motivations for this proposal were explained there.  Another idea
which originally motivated that equation is the following: When one carries
out the constructions of field theory with these new local objects---objects
in~$\scrK^{\bullet}(X)$---the geometric meaning of certain quantities
changes.  In particular, the Noether currents associated to gauge symmetries
will now also be elements of~$\scrK^{\bullet}(X)$, but with opposite parity.
This makes it natural that the Noether charge, the topological equivalence
class of the Noether current, is an element of the appropriate $K$-theory
group, as in~\cite{W4}.

As mentioned, there is a geometric version of any generalized cohomology
theory.  We will make use of this for $KO$-theory (real bundles) and
$KSp$-theory (quaternionic bundles).  We term such theories {\it differential
$K$-theory\/}, {\it differential $KO$-theory\/}, etc.

 \subhead Pfaffians of Dirac operators
 \endsubhead

This is well-known material.
 
The functional integral over a free spinor field is formally the pfaffian of
a Dirac operator, which is an element of a real or complex line.  If $W\to
S$~is a family of $n$-dimensional spin manifolds, then the pfaffian is a
section of a real or complex line bundle with metric and connection over~$S$.
We consider Dirac operators with coefficients in a bundle~$E\to W$, which may
be real, complex, or quaternionic.  Now up to equivalence a real line bundle
with metric and connection over~$S$ is an element of~$H^1(S;\zt)$; in the
complex case it is an element of~$\HH^2(S)$.  The construction of these
bundles depends on~$n\pmod8$, and we quickly review the constructions in the
language of the previous section.  (For a description of the even dimensional
case, see~\cite{F2}; for the odd dimensional case, see~\cite{S} and
also~\cite{MW,\S4.1}.)

A geometric form of the index theorem asserts that the index of the family of
Dirac operators is the image of~$E$ under integration (pushforward) in
differential $K$-theory, which is a map
  $$ \KK(W)\longrightarrow \KK^{-n}(S). \tag{13} $$
We then compose with a determinant or pfaffian to obtain the appropriate line
bundle.  The map~\thetag{13} has refinements in real and symplectic
$K$-theory, and we need these to obtain pfaffians. 
 
The even dimensional case goes as follows.  For $n\equiv 0,4\pmod8$ we may as
well suppose the bundle~$E$ is complex, in which case by periodicity the
image of~$E$ under~\thetag{13} lies in~$\KK(S)$.  The {\it determinant\/} line
bundle is obtained by a map $\KK(S)\to \HH^2(S)$ which refines the usual
topological determinant line bundle in $K$-theory.  In quantum field theories
CPT~invariance dictates that if there are positive half-spinors with values
in~$E$, then there are also negative half-spinors with values in~$\Ebar$.
Formally the fermionic path integral is the {\it pfaffian\/} of~$D(E-\Ebar)$,
Dirac coupled to the formal difference~$E-\Ebar$.  But the determinant line
bundle~$\Det D(E-\Ebar)$ is isomorphic to~$\Det D(E)^{\otimes 2}$, so we
take~$\Pfaff D(E-\Ebar)$ to be~$\Det D(E)$.  For $n\equiv 2\pmod 8$ we obtain
a pfaffian if $E$~is real.  Then the index lies in~$\KO^{-2}(S)$, and there
is a pfaffian line bundle $\KO^{-2}(S)\to\HH^2(S)$.  Similarly, for $n\equiv
6\pmod8$ if we start with a quaternionic bundle~$E$, then we apply~\thetag{13}
to $\KSp$ thereby obtaining an index in~$\KSp^{-6}(S)\cong \KO^{-2}(S)$ and
so a pfaffian as before.
 
In the odd dimensional case the complex Dirac operator is self-adjoint, but
the determinant is naturally a complex number---the exponentiated $\eta
$-invariant enters.  The square root is then a section of a real line bundle,
and the construction of this bundle is topological.  For $n\equiv 1\pmod 8$
we start with a real bundle~$E$.  The index lies in~$KO^{-1}(S)$, and the
pfaffian line bundle is obtained by the natural map $KO^{-1}(S)\to
H^1(S;\zt)$.  For $n\equiv 3\pmod8$ we also start with a real bundle, and
again there is a natural map $KO^{-3}(S)\to H^1(S;\zt)$ (which factors
through a map $KO^{-3}(S)\to H^1(S;\ZZ)$).  For $n\equiv 5,7\pmod8$ we take
$E$~to be quaternionic, and by periodicity we obtain the same constructions
of the pfaffian.

 \subhead Coupling to D-branes
 \endsubhead

The spacetime~$X$ of Type~II theory is a spin Riemannian 10-manifold, which
for simplicity we assume to be compact.  The worldvolume~$W$ of a D-brane is
a submanifold endowed with a complex vector bundle~$Q\to W$.  There are also
scalar fields and fermions on the brane.  In the basic case this Chan-Paton
bundle~$Q$ has rank one, corresponding to unit charge, and as demonstrated
in~\cite{FW} is more properly viewed as a spin$^c$ structure on~$W$.  For
simplicity we assume that $W$~is spin, and allow for arbitrary rank vector
bundles~$Q$.  In fact, $Q$~comes equipped with differential geometric data
and should be viewed as an element of~$\scrK^0(W)$.
 
Finally we are ready to describe the coupling of the D-brane to the
RR-field~$C\in \scrK^q(X)$.  Since the field strength of~$C$ is self-dual,
the D-brane is both electrically and magnetically charged under~$C$.  The
magnetic charge means that the geometric meaning of~$C$ is shifted.  Namely,
the D-brane charge is represented by the pushforward of~$Q$ to the bulk~$X$,
which is an element of~$\scrK^p(X)$, where $p$~is the codimension of~$W$
in~$X$.  As explained earlier, the precise construction of a {\it smooth\/}
element depends on some choices.  Then $C$~is a trivialization of this
element which obeys the following constraint.  The restriction of the D-brane
charge to~$W$ is $Q$ times the Euler class of the normal bundle to~$W$ in
$K$-theory.  If $p\equiv -1,0,1\pmod 8$, then this Euler class has a real
refinement, and we constrain $\Qbar\,C$ to be a trivialization of ~$\Qbar\,Q$
times this real refinement.  If $p\equiv 3,4,5\pmod8$, then this Euler class
has a quaternionic structure, and we constrain $\Qbar\,C$ to be
compatible.\footnote{These statements requires a technical explanation,
details of which will appear in our subsequent paper.  Let $\nu \to W$ be the
normal bundle, which is real of rank~$p$.  If $p$~is even, then the
$K$-theory Euler class in~$K(W)$ may be identified with the difference of the
half-spinor bundles of~$\nu $.  It has a real structure if~$p\equiv 0\pmod8$,
a quaternionic structure if~$p\equiv 4\pmod 8$, and is complex (not
self-conjugate) if $p\equiv 2,6\pmod8$.  In these cases we simply regard the
complex $K$-theory Euler class as an element of the appropriate differential
$KO$-, $KSp$-, or $K$-group.  If $p$~is odd, the Euler class in topological
$K$-theory vanishes, but in differential $K$-theory it is an element of order~2
in $K^0(X;\RR)/K^0(X)\hookrightarrow \KK^1(X)$.  It can be identified with
the mod~2 reduction of the spinor bundle~$S(\nu )$.  We summarize the Euler
class computation in complex differential $K$-theory by the equation 
  $$ \Euler_{\KK}(\nu ) = \cases S^+(\nu ) - S^-(\nu ) ,&\text{rank($\nu $)
     even};\\S(\nu )\pmod2,&\text{rank($\nu $) odd}.\endcases \tag{14} $$
} As we explained after
equation~\thetag{6} there is an extra factor of~$1/2$ in the electric and
magnetic coupling of a self-dual field.  The constraint just described is our
interpretation of this factor in the magnetic charge.  The electric coupling
is a term in the action:
  $$ i\int_{W} \frac 12\,\Qbar \,C. \tag{15} $$
The interpretation of this term depends on the dimension.  For $p\equiv
-1,0,1\pmod8$ we interpret the exponential of~\thetag{15} as $\exp\left(
i\int_{W}\Qbar\,C \right) $ in $\KO$-theory.  The cases $p\equiv 3,4,5\pmod8$
are similar, except that we use $\KSp$-theory.  For $p\equiv 2,6\pmod8$ the
square of the exponential of~\thetag{15}, computed in $\KK$-theory, is a
section of a line bundle which is a square, as explained in the previous
section.  We take~\thetag{15} to be one of the square roots of that section.

As in~\thetag{7} this term is of Green-Schwarz type---its exponential is a
section of a line bundle with connection over the space of parameters.  Thus
it contributes to the overall anomaly.  Let $W\to S$ be a family of D-branes
parametrized by a manifold~$S$.  This comes as a fiberwise submanifold of a
family $X\to S$ of 10-dimensional spacetimes, which we do not require be a
product.  The line bundle in question is computed by
integrating~$\Qbar\,Q\cdot \operatorname{Spinors}(\nu )$ over the fibers
of~$W\to S$ in the appropriate differential $K$-theory.  As explained in the
previous paragraph, we identify the differential $K$-theory Euler class with
``$\operatorname{Spinors}(\nu)$'', which if $p$~is odd is identified with the
spin bundle of~$\nu $ modulo two and if $p$~is even is the difference of
half-spin bundles.  In all cases it is to be regarded in the appropriate
$\KK$-, $\KO$-, or $\KSp$-group.  (For $p\equiv 2,6\pmod8$ we use the square
root mentioned at the end of the previous paragraph.)
 
The (complex) fermi field on~$W$ is the restriction of a chiral spinor in~$X$
to~$W$ tensored with~$\End(Q)\cong \Qbar\,Q$.  Since we assume $W$~is spin,
the restriction to~$W$ of the bundle of half-spinors on~$X$ decomposes.  If
$p$~is odd we obtain the tensor product of spin bundles~$S(W)\otimes S(\nu
)$, whereas if $p$~is even we have the sum $S^+(W)\otimes S^+(\nu )\,\oplus
\,S^-(W)\otimes S^-(\nu )$.  The fermionic functional integral---Dirac
pfaffian---is computed by coupling Dirac to~$S(\nu )$ in the odd case and the
formal difference $S^+(\nu )- S^-(\nu )$ in the even case.  As explained in
the previous section, the pfaffian line bundle is computed by the
Atiyah-Singer index theorem: we regard these spin bundles in the appropriate
differential complex, real, or quaternionic $K$-group and integrate.
Therefore, we get precisely the same line bundle with connection as in the
previous paragraph.  Presuming that the coupling term~\thetag{15} comes with
an overall minus sign, we see that the anomaly from the fermions on the brane
cancels the anomaly from~\thetag{15}.

\enddocument